\documentclass[english,aps,pra,reprint,noshowpacs,superscriptaddress]{revtex4-1}   
\usepackage{geometry,graphicx,times,hyperref,enumitem,amsmath}
\usepackage[T1]{fontenc}
\usepackage[latin9]{inputenc}
\usepackage[above,below]{placeins}
\usepackage[dvipsnames]{xcolor}
\hypersetup{colorlinks=true,urlcolor=blue,citecolor=blue,linkcolor=blue}   
\newcommand{\drdf}[1]{#1}
\newcommand{\rios}[1]{#1}
\newcommand{\quan}[1]{#1}

\newcommand{\FIGcorrelations}{
\begin{figure*}[t]
\includegraphics[width=\textwidth]{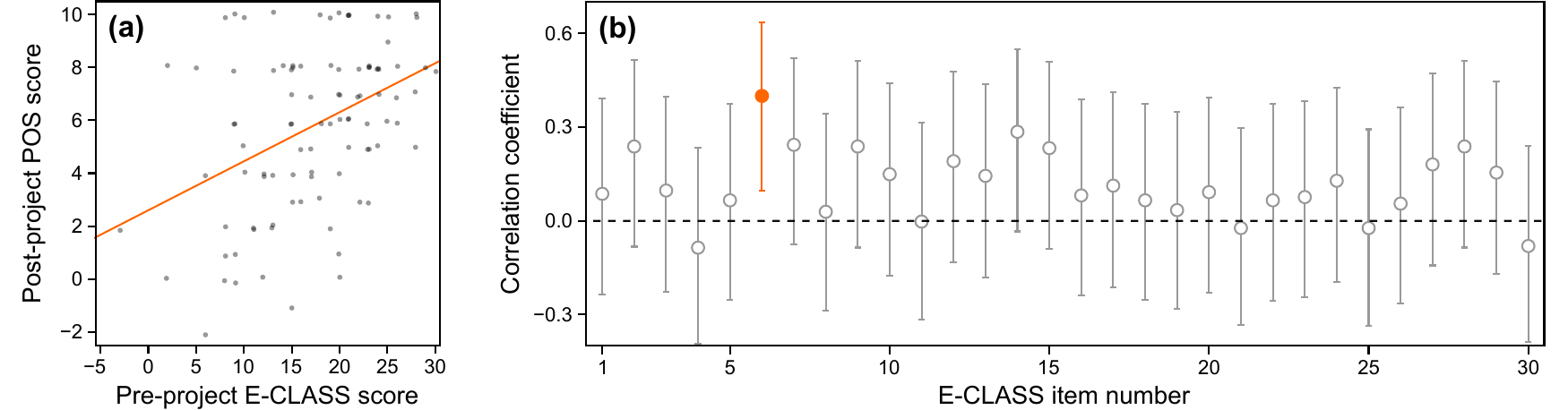}
\caption{\label{fig:correlations}(a) Scatter plot of matched POS and E-CLASS scores. To improve visibility of overlapping markers, a small artificial jitter has been added to each data point, and overlapping markers are darker than nonoverlapping ones. The curve is a line of best fit representing the positive correlation between POS and E-CLASS scores ($r=0.41$, $p\ll0.001$). (b) Correlations between POS score and each individual E-CLASS item score. Circular markers represent correlation coefficients. Error bars correspond to 95\% confidence intervals, corrected for multiple comparisons. Only item 6 yielded a statistically significant correlation ($r=0.40$, $p\ll0.001$), indicated by a filled marker.}
\end{figure*}
}

\newcommand{\TABdemo}{
\begin{table}[b]
\caption{\label{tab:demo}Courses in our study. $N$ is the number of students for whom we have matched E-CLASS and POS scores. We collected data for two instances of \textdagger\ and one instance of all other courses.}
\begin{ruledtabular}
\begin{tabular}{llcr}
Institution & Course & Project weeks & $N$ \\ \hline
Selective, public, doctoral & Advanced Lab & 5 & 8\\
Inclusive, public, master's & Advanced Lab & 4 & 8\\
Inclusive, public, master's & Advanced Lab & 4 & 13\\
Selective, private, bac. & Experimental Phys. & 4 &15\\ 
Selective, private, master's & Lasers Lab & 7 &19\\
& Optics Lab\textsuperscript{\textdagger} & 7 & 33 \\
\end{tabular}
\end{ruledtabular}
\end{table}
}

\begin{document}

\title{Correlating students' views about experimental physics with their sense of project ownership}
\author{Dimitri R. Dounas-Frazer}
\affiliation{Department of Physics, University of Colorado Boulder, Boulder, CO 80309, USA}
\affiliation{JILA, National Institute of Standards and Technology and University of Colorado Boulder, Boulder, CO 80309, USA}
\author{H. J. Lewandowski}
\affiliation{Department of Physics, University of Colorado Boulder, Boulder, CO 80309, USA}
\affiliation{JILA, National Institute of Standards and Technology and University of Colorado Boulder, Boulder, CO 80309, USA}


\begin{abstract}
\drdf{Multiweek projects in physics labs can engage students in authentic experimentation practices}, and it is important to understand student experiences \rios{during projects} along multiple dimensions. To this end, we conducted an exploratory quantitative investigation to look for connections between students' pre-project views about experimental physics and their post-project sense of project ownership. We administered the Colorado Learning Attitudes About Science Survey for Experimental Physics (E-CLASS) and the Project Ownership Survey (POS) to 96 students enrolled in 6 lab courses at 5 universities. E-CLASS and POS scores were positively correlated, suggesting that students' views about experimentation may be linked to their ownership of projects. This finding motivates future studies that could explore whether these constructs are causally related.
\end{abstract}

\maketitle


\section{Introduction}
\vspace{-5pt}

Project-based learning is becoming increasingly popular in undergraduate physics education~\cite{Feder2017}. Meanwhile, physics educators and education researchers are investing in the improvement of instructional lab courses~\cite{Caballero2018}. In labs, multiweek projects are a promising way to engage undergraduate students in authentic experimental physics practices. Across lab course contexts---and sometimes even within a single course---project topics span a wide range of physics concepts and require various types of equipment. Nevertheless, some features of \rios{project experiences} are decoupled from particular phenomena and apparatus. Projects can provide opportunities for students to take control of an experiment, collaborate with peers or instructors to brainstorm solutions to problems, implement and revise their own ideas, and experience the joys and frustrations of experimentation~\cite{Dounas-Frazer2017}. Research on these and other cognitive, social, or affective aspects of projects in undergraduate physics labs is sparse, pointing to a need for exploratory and theory-building investigations of students' experiences while working on projects.

In this paper, we explore connections between students' views about experimental physics and their sense of project ownership. Views about experimental physics include students' ideas about which skills, practices, or goals are important for conducting experiments, as well as their beliefs about whether experimentation is something that they can do or that they enjoy~\cite{Zwickl2014}. Sense of project ownership arises when students have the right and responsibility to make their own decisions about the project, care about the outcome of the project, or feel a personal connection to the project and identify it as ``their own"~\cite{Wiley2009}. \drdf{Based on our experience working with lab instructors, one common goal of projects is for students to feel ownership of their experiments. To facilitate this goal, we ultimately aim to develop theories about which factors support students' development of project ownership in lab courses.} We suspect that understanding the nature of experimentation may be one such factor.
\quan{Because identifying correlations can help motivate more intensive searches for causal links,}
we ask: when working on multiweek projects, do students' pre-project views about experimental physics correlate with their post-project sense of project ownership?

To answer our research question, we conducted a cross-institutional exploratory quantitative study using two  \drdf{established} survey instruments: the Colorado Learning Attitudes About Science Survey for Experimental Physics (E-CLASS)~\cite{Wilcox2016b}, and the Project Ownership Survey (POS)~\cite{Hanauer2014}. We used \drdf{the} E-CLASS and POS to measure students' pre-project views about experimental physics and their post-project sense of project ownership, respectively. Based on 96 matched survey responses, we found a statistically significant moderate positive correlation between students' E-CLASS and POS scores. \rios{In the next section, we summarize relevant literature to provide context for this finding.} 


\vspace{-5pt}
\section{Background and motivation}

\rios{In addition to summarizing prior research using E-CLASS and POS, we propose mechanisms that could link students' views about experimental physics and their sense of project ownership---and hence their E-CLASS and POS scores.}


\vspace{-5pt}
\subsection{\label{sec:E-CLASS}\rios{Prior research using E-CLASS}}

Most previous investigations of students' views about experimental physics are large quantitative studies that use E-CLASS data collected in lab courses from across the country. Developed by Zwickl et al.~\cite{Zwickl2014}, E-CLASS is a Likert-style survey with 30 items. Each item is a statement about experimental physics with which the student is prompted to agree or disagree using a 5-point scale. Example statements include, ``Designing and building things is an important part of doing physics experiments," and, ``Communicating scientific results to peers is a valuable part of doing physics experiments." For each item, a student's response is scored as expertlike if it aligns with experts' responses to that item~\cite{Zwickl2014}.

\quan{In a national study, Wilcox and Lewandowski~\cite{Wilcox2017c}}
\rios{showed that, on average, students in first-year labs had expertlike responses to about two-thirds of E-CLASS items.
Students in higher level labs, on the other hand, had expertlike responses to about three-quarters of items.}
Regardless of course level, fewer than half of students had expertlike responses to items ``related to what can be loosely described as student autonomy, or their ability to direct an experiment, overcome difficulties, and select analysis methods without guidance from an authority figure."~\cite{Wilcox2017c} (p.~6).

To provide additional insight into students' views about experimental physics, Hu et al.~\cite{Hu2017} modified E-CLASS by adding free-response questions to a few key survey items. For these items, free-response questions probed students' rationale for agreeing or disagreeing with a particular statement. Qualitative analysis of over 200 modified surveys revealed several patterns in student reasoning. One common line of reasoning was the following: experiments whose goal is to confirm previously known results may nevertheless contribute to the growth of scientific knowledge \drdf{by enhancing students'
own \emph{personal} scientific knowledge~\cite{Hu2017}.}

\rios{This previous work~\cite{Wilcox2017c,Hu2017} demonstrates that E-CLASS is sensitive to students' beliefs about their own autonomy and the growth of their own personal knowledge.}
Such beliefs are related to aspects of ownership, namely, being responsible for an experiment and feeling personally connected to its outcome~\cite{Wiley2009}. Before drawing these connections more explicitly, we first summarize relevant research on student ownership.


\vspace{-5pt}
\subsection{\label{sec:POS}\rios{Prior research using POS}}

Previous investigations of students' sense of project ownership include quantitative studies that use POS data collected in biology or physics lab courses. Developed by Hanauer and Dolan~\cite{Hanauer2014}, POS is a Likert-style survey with 16 items that can be grouped into two parts. The first part consists of 10 items. Each of these items is a statement about research projects with which the student is prompted to agree or disagree using a 5-point scale. Example statements include, ``I was responsible for the outcome of my research," and, ``The findings of my research project gave me a sense of personal achievement." Agreement is indicative of \rios{project ownership}. The second part of POS consists of 6 items. Each of these items is a question of the form, ``To what extent does the word \emph{happy} describe your experience in the laboratory course?" All of the emotive questions probe positive feelings about the course. Possible responses range from ``very strongly" to ``very slightly" on a 5-point scale. Strong associations with positive feelings about the course were intended to be indicative of a sense of project ownership~\cite{Hanauer2014}, but more recent research calls the interpretation of the emotive POS items into question in physics lab contexts~\cite{Stanley2016}.

\quan{In a national study, Hanauer and Dolan~\cite{Hanauer2014}}
showed that, on average, students in research-based lab courses had more positive scores on all POS items than their counterparts in traditional labs. On 11 items, the differences were statistically significant with medium-to-large effect sizes. The largest differences corresponded to statements related to students' sense of personal achievement, solicitation of advice and assistance, and belief that their findings were important to the scientific community. The emotive questions related to happiness, delight, and joy also discriminated between research-based and traditional labs~\cite{Hanauer2014}. However, subsequent implementations of POS in a physics context suggest that the emotive questions are not always good indicators of project ownership~\cite{Stanley2016}.

Stanley et al.~\cite{Stanley2016} implemented POS in an optics lab with multiweek projects. Post-project interviews indicated that almost all students had high levels of ownership. Overall, 36 students completed POS over two semesters, yielding positive scores on the first 10 survey items but neutral scores on the 6 emotive questions. Interview data suggested that students had complex emotional experiences while working on their projects, and POS likely failed to capture this complexity~\cite{Stanley2016}. In a follow-up investigation, Dounas-Frazer et al.~\cite{Dounas-Frazer2017} conducted a multiple case study of three student groups from the optics lab. \drdf{They found that ``student ownership is characterized by a wide range of emotions that fluctuate in time as students alternate between extended periods of struggle and moments of success while working on their project." (p.~19).} This finding points to a possible explanation for why emotive questions on POS are difficult to interpret: when asked about the extent to which a particular emotion describes their experience, some students may respond by ``integrating" their emotional fluctuations over the duration of the project while others may simply indicate how they feel at the time the survey is administered, i.e., at the end of the project~\cite{Stanley2016}.


\vspace{-5pt}
\subsection{\label{sec:connections}Connections between views about experimental physics\\ and sense of project ownership}

It is plausible that, when working on a project, students' views about experimental physics can promote their sense of project ownership. For example, Quan and Elby~\cite{Quan2016} explored the coupling of students' views about the nature of science and their self-efficacy toward physics research. They demonstrated that, for some undergraduate students, ``shifts in seeing research as a place where novices can participate led students to see themselves as able to make a meaningful research contribution." (p.~10). Self-efficacy is related to, but different from, project ownership~\cite{Dounas-Frazer2017}. Nevertheless, perceiving one's own contributions to research as meaningful is similar to feeling a sense of personal achievement when working on a project, which is a component of project ownership~\cite{Wiley2009,Dounas-Frazer2017,Hanauer2014}. \quan{Inspired by Quan and Elby~\cite{Quan2016} and Hu et al.~\cite{Hu2017}, we speculate that students'  sense of project ownership may be coupled to their view of experimentation as a process that they can meaningfully participate in and learn from.}

\FIGcorrelations

We can further imagine another mechanism through which views about experimental physics may be connected to project ownership. If students believe that experimentation involves encountering and overcoming difficulties, they may be more likely to attribute slow progress or unanticipated obstacles to the nature of experimentation rather than their own abilities. In turn, this may positively impact their ability to cope with frustration and tedium while troubleshooting problems, ultimately contributing to a sense of personal achievement when they solve problems on their own (cf. Ref.~\cite{Dounas-Frazer2017}). Indeed, we have previously argued that a belief that ``nothing works the first time" is an expertlike epistemology that can impact how students design and build apparatus~\cite{Dounas-Frazer2016b}.

\drdf{Thus, multiple hypothetical mechanisms could link views about experimentation and sense of ownership. Here, we use E-CLASS and POS to look for evidence of such a link.}


\vspace{-5pt}
\section{\label{sec:methods}Methods}

We administered E-CLASS and POS to students in 6 lab courses at 5 institutions. Each course included a multiweek project, ranging in duration from 4 to 7 weeks. Students worked in groups of 2 to 4, and projects culminated in final reports and oral presentations. Project topics included achieving supersonic levitation of small objects, measuring the elastic modulus of a piezoelectric material, and generating images of objects hidden behind a scattering material.
\drdf{In multiple courses, students used journal articles to help propose project topics, design apparatus, analyze data, evaluate progress, or solve problems that arose.}
E-CLASS was administered at the start of the course and POS at the end, i.e., pre- and post-project, respectively. A total of 96 students completed both surveys, corresponding to a response rate of 75\%. A breakdown of participation by course is provided in Table~\ref{tab:demo}.

\TABdemo

For E-CLASS, we analyzed all 30 items that probe students' personal views about experimental physics. For POS, we interpret the two parts of the survey as measuring distinct aspects of students' sense of project ownership, as do others~\cite{Stanley2016,Hanauer2016}. Because interpretation of responses to emotive questions is unclear~\cite{Stanley2016}, we \drdf{included only} the first 10 survey items in our analysis. For both E-CLASS and POS, we collapsed item scores from a 5-point to a 3-point scale: unfavorable ($-1$), neutral ($0$), or favorable ($+1$). Here, expertlike E-CLASS responses and positive POS responses are ``favorable." The total score for each survey is a sum of all the item scores. E-CLASS scores range from $-30$ to $+30$, with negative and positive scores corresponding to beliefs that are inconsistent or consistent with experts. Similarly, POS scores range from $-10$ to $+10$, with negative and positive scores indicating the absence or presence of ownership.

For each student who completed both surveys, we matched pre-project E-CLASS and post-project POS responses. Next, we computed the correlation between overall E-CLASS and POS scores. Whereas POS was designed to probe a single construct~\cite{Hanauer2014}, E-CLASS was not~\cite{Wilcox2016b}. Therefore, we also computed correlations between each individual E-CLASS item and POS. For a given correlation, we computed a Pearson correlation coefficient, $r$. \quan{This coefficient ranges from $-1$ to $+1$, with larger magnitudes corresponding to stronger correlations.} Confidence intervals and $p$-values \quan{for correlation coefficients} were determined using a standard Fisher transformation and a two-tailed t-test, respectively. For correlations between individual E-CLASS items and POS, we incorporated a Bonferroni correction for multiple comparisons~\cite{Agresti2009}.


\vspace{-5pt}
\section{\label{sec:results}Results and discussion}

A scatterplot of matched E-CLASS and POS scores is shown in Fig.~\ref{fig:correlations}(a). We found a moderate positive correlation between E-CLASS and POS: $r=0.41$ with 95\% confidence interval $[0.23,0.57]$ and $p\ll 0.001$. Correlation coefficients for individual E-CLASS items are shown in Fig.~\ref{fig:correlations}(b). Most itemwise correlation coefficients were positive and a few were negative, but only one was statistically different from zero. Moreover, the 95\% confidence intervals for itemwise correlations all overlapped with each other, indicating that no single E-CLASS item was significantly different from the others. Therefore, we do not see evidence that the the moderate positive correlation between overall E-CLASS and POS scores was driven by a subset of E-CLASS items. Rather, the survey as a whole correlates with POS.

The only E-CLASS item that had a statistically significant correlation with POS was item 6: ``Scientific journal articles are helpful for answering my own questions and designing experiments." For this item, the correlation was moderate and positive: $r=0.40$ with 95\% confidence interval $[0.10,0.63]$ and $p\ll0.001$. This result is understandable given that multiple courses in our study required students to use journal articles throughout their projects. Students with more expertlike views about the role of journal articles in experimentation may have been more willing or likely to use scientific literature when appropriate, thus facilitating their autonomy and ownership with respect to the project. However, while item 6 was the only item with a nonzero correlation, we emphasize that the overall correlation between E-CLASS and POS was not due to this item alone.

The observed positive correlation between pre-project E-CLASS and post-project POS scores demonstrates that students' views about experimental physics and their sense of project ownership are coupled, and it raises questions about why this is the case. Perhaps, as we suggested in Sec.~\ref{sec:connections}, there are causal mechanisms through which students' expertlike views about experimentation result in improved self-efficacy, anticipation of unexpected problems, or emotional regulation (cf. Refs.~\cite{Quan2016,Dounas-Frazer2016b,Dounas-Frazer2017}), which position students to propose and implement their own ideas when working on projects. \rios{Another explanation could be that students' previous experience working on projects simultaneously shifted their beliefs about what experimentation entails and increased their interest in articulating and exploring their own questions.} Previous project experience could thus be a confounding variable that gives rise to non-causal correlations between E-CLASS and POS scores. \quan{Further exploratory research, such as qualitative interview studies (cf. Ref.~\cite{Quan2016}), is needed to illustrate mechanisms that lead to the positive correlations we have observed.}


\vspace{-5pt}
\section{\label{sec:conclusion}Future directions}

In the future, we aim to build on the results presented here using both quantitative and qualitative methods. We plan to continue implementing E-CLASS and POS in the same six lab courses each spring. By increasing the number of students in our study, our analyses may become sensitive to inter-course differences or allow for exploring impacts of potential confounders (e.g., students' prior project experience). We may also be able to distinguish some individual E-CLASS items from others in terms of itemwise correlations with POS. 

In anticipation of future qualitative work, we have conducted post-project interviews with students and instructors in each of the courses described here. Additionally, we have collected project artifacts, including weekly free-response surveys about students' experiences. These data may help us understand which mechanisms give rise to coupling of students' views about experimental physics with their sense of project ownership. \drdf{Identifying such mechanisms is an important step toward developing effective pedagogies for physics lab courses that prioritize students' sense of project ownership as a learning outcome.}


\vspace{-5pt}
\begin{acknowledgments}
We thank Gina M. Quan and Laura R\'ios for their valuable input. This material is based upon work supported by the NSF under Grant Nos. DUE-1726045 and PHY-1734006.
\end{acknowledgments}


\bibliography{./perc2018_database}

\end{document}